\documentclass[pdflatex,sn-mathphys-num]{sn-jnl}

\usepackage{graphicx}%
\usepackage{multirow}%
\usepackage{amsmath,amssymb,amsfonts}%
\usepackage{amsthm}%
\usepackage{mathrsfs}%
\usepackage[title]{appendix}%
\usepackage{xcolor}%
\usepackage{textcomp}%
\usepackage{manyfoot}%
\usepackage{booktabs}%
\usepackage{algorithm}%
\usepackage{algorithmicx}%
\usepackage{algpseudocode}%
\usepackage{listings}%
\usepackage{verbatim}
\usepackage{dirtytalk}

\usepackage[backend=biber, style=numeric, sorting=none]{biblatex}
\begin{document}

\title[Article Title]{Three-dimensional excitonic dipole anisotropy enables ultrabroadband polarization photodetection in CrCl$_3$}

\author[1]{\fnm{Satyam} \sur{Sahu}}\email{satyam.sahu@jh-inst.cas.cz}
\author[1,2]{\fnm{Jaganandha} \sur{Panda}}\email{jagan.panda@uni-muenster.de}
\author[1]{\fnm{Martin} \sur{Jindra}}\email{martin.jindra@jh-inst.cas.cz}
\author[1,3]{\fnm{Mukesh} \sur{Kumar Thakur}}\email{mukesh@iiitvadodara.ac.in}
\author[1,4]{\fnm{Farjana J.} \sur{Sonia}}\email{f.j.sonia@ifw-dresden.de}
\author[5]{\fnm{Shankar} \sur{Khanal}}\email{skhanal@mag.mff.cuni.cz}
\author[4]{\fnm{Kornelius} \sur{Nielsch}}\email{k.nielsch@ifw-dresden.de}
\author[5]{\fnm{Jana} \sur{Vejpravova}}\email{jana.vejpravova@matfyz.cuni.cz}
\author[1]{\fnm{Matěj} \sur{Velický}}\email{matej.velicky@jh-inst.cas.cz}
\author*[1]{\fnm{Martin} \sur{Kalbáč}}\email{martin.kalbac@jh-inst.cas.cz}
\author*[1]{\fnm{Otakar} \sur{Frank}}\email{otakar.frank@jh-inst.cas.cz}
\author*[1,4]{\fnm{Golam} \sur{Haider}}\email{g.haider@ifw-dresden.de}

\affil*[1]{\orgname{Heyrovský Institute of the Czech Academy of Sciences}, \orgaddress{\street{Dolejškova 2155/3}, \city{Prague}, \postcode{18200}, \country{Czech Republic}}}

\affil[2]{\orgname{Münster Nanofabrication Facility, University of Münster}, \orgaddress{\street{Busso-Peus-Str.10}, \city{Münster}, \postcode{48149}, \country{Germany}}}


\affil[3]{\orgname{Department of Applied Physics, Indian Institute of Information Technology Vadodara}, \orgaddress{\city{Gandhinagar}, \postcode{382028}, \country{India}}}

\affil*[4]{\orgname{Leibniz Institute for Solid State and Materials Research}, \orgaddress{\street{Helmholtzstr. 20}, \city{Dresden}, \postcode{01069}, \country{Germany}}}

\affil[5]{\orgname{Department of Condensed Matter Physics, Charles University}, \orgaddress{\street{Ke Karlovu 5}, \city{Prague}, \postcode{12116}, \country{Czech Republic}}}
\abstract{Simultaneous detection of the spectral and polarization properties of light is highly desirable for integrated imaging and photonic technologies but typically requires complex multi-component architectures. Here, we demonstrate that the intrinsic dielectric anisotropy of layered insulating CrCl$_3$ enables ultrabroadband polarization-resolved photodetection spanning wavelengths from 300 to 1700 nm. The photoresponse is governed by long-lived ligand-field excitons, whose microsecond-scale lifetime produces a photoconductive gain exceeding $4.5 \times 10^{4}$. By combining wavelength-, polarization-, and angle-resolved optoelectronic measurements, we reveal that distinct ligand-field and higher-energy excitonic transitions possess different optical dipole orientations, leading to excitation-energy-dependent rotation of the in-plane polarization axis. Furthermore, oblique illumination activates out-of-plane optical dipoles, while competing excitonic transitions with distinct dipole orientations drive wavelength-dependent rotation and reversal of the polarization anisotropy. Together, these effects produce a highly tunable degree of polarization ranging from $-$90\% to $+$75\%, establishing intrinsic three-dimensional vectorial light-matter interactions in a layered magnetic van der Waals insulator. These findings establish dielectric anisotropy and excitonic dipole engineering as powerful design principles for compact ultrabroadband polarization-sensitive photodetectors and multifunctional van der Waals photonic systems.}


\maketitle
\section{Introduction}\label{intro}
The quest for advanced optoelectronic devices has been driven by the need to simultaneously achieve broad spectral coverage, high sensitivity, and precise polarization control. Photodetectors capable of addressing these requirements are essential for applications ranging from optical communication and remote sensing to quantum information processing and biomedical imaging \cite{Du2024, Liu2025, Pan2025}. Conventional semiconductor platforms have delivered remarkable progress, yet their operation is often constrained by limited spectral ranges and weak anisotropic responses, thereby motivating the exploration of new materials systems with unconventional light-matter interactions \cite{Liu2025, Ma2024, Chen2025, Li2025, Zhang2025, Su2025}.\par
Two-dimensional (2D) van der Waals (vdW) materials have opened exciting possibilities in this regard. Their atomic thinness, tunable band structures, and ability to integrate seamlessly with diverse platforms make them attractive for next-generation photodetectors \cite{Koppens2014, Lopez2013, Xia2014}. Materials such as graphene, transition metal dichalcogenides, and black phosphorus have demonstrated distinctive photoresponse mechanisms, including ultrafast carrier dynamics and anisotropic optical absorption \cite{Xia2009, Low2014, Qiao2014}. However, despite this rapid progress, simultaneously realizing ultrabroadband photodetection with robust polarization sensitivity in both the in-plane and out-of-plane directions has remained a formidable challenge. Most 2D photodetectors reported to date either lack a sufficiently broadband response or are restricted to in-plane polarization selectivity, leaving a significant gap for multifunctional devices that combine these properties in a single platform \cite{Su2025, Yu2025, Huang2025}. Recent advances in anisotropic vdW materials have further expanded the scope of optoelectronics by enabling functionalities such as helicity-sensitive \cite{Cheng2021} and anisotropic photodetection \cite{Panda2023, Zhou2025, Guo2022}, highlighting the potential of layered anisotropic materials as multifunctional photoactive platforms. Despite these developments, the optoelectronic response of many such systems remains largely unexplored, particularly in relation to polarization-dependent light-matter interactions.\par
Among them, CrCl$_3$ has attracted considerable attention owing to its ligand-field excitons and rich optical response arising from localized Cr$^{3+}$ electronic states. At room temperature, bulk CrCl$_3$ crystallizes in a monoclinic structure (space group $C2/m$) composed of weakly coupled vdW layers formed by edge-sharing CrCl$_6$ octahedra (Figure \ref{fig:Raman_PL}a). Furthermore, bulk CrCl$_3$ flakes exhibit good ambient stability, making CrCl$_3$ an attractive platform for optical and spectroscopic investigations under standard conditions \cite{Wang2023}. Unlike many semiconducting vdW materials whose optical response is dominated by band-to-band transitions, the optical properties of CrCl$_3$ are strongly influenced by localized electronic states associated with the Cr$^{3+}$ ions. In addition to its wide band gap of approximately 3.3 eV \cite{Ermolaev2025}, CrCl$_3$ exhibits characteristic sub-gap optical absorption and emission arising from crystal-field (ligand-field; d-d) transitions within the partially filled 3d shell of Cr$^{3+}$ \cite{Ermolaev2025, Polini1970, Cai2019}. Recent studies have further shown that these strongly bound ligand-field excitons play an important role in the optoelectronic response of bulk CrCl$_3$ and exhibit long-lived relaxation dynamics \cite{Sridhar2024}. These transitions occur at energies well below the fundamental band gap and can contribute significantly to the optical response throughout the visible (Vis) and near-infrared (NIR) spectral range. Owing to their localized nature and sensitivity to the crystal environment, ligand-field transitions provide a unique platform for exploring anisotropic light-matter interactions in low-symmetry vdW crystals.\par
In this work, we show that the intrinsic dielectric anisotropy of layered CrCl$_3$ governs a three-dimensional excitonic dipole landscape that enables simultaneous ultrabroadband photodetection and polarization selectivity. By combining wavelength-, polarization-, and angle-resolved optoelectronic measurements with polarization-resolved emission and reflectivity spectroscopy, we demonstrate that distinct ligand-field and higher-energy excitonic transitions possess different optical dipole orientations, leading to excitation-energy-dependent rotation of the polarization axis and the emergence of out-of-plane optical dipoles. These results establish dielectric anisotropy as a fundamental mechanism for engineering vectorial light-matter interactions in magnetic van der Waals semiconductors and provide a design strategy for compact multifunctional photonic devices.\par

\section{Results and discussion}
\subsection{Optical characterization}\label{result_char}
We begin by establishing the structural and optical fingerprints of chemical vapor transport grown CrCl$_3$ bulk flakes (see Experimental Section for details) mechanically exfoliated on SiO$_2$/Si using Raman and photoluminescence (PL) spectroscopy.\par
\begin{figure}[htbp!]
    \centering
    \includegraphics[width=0.90\textwidth]{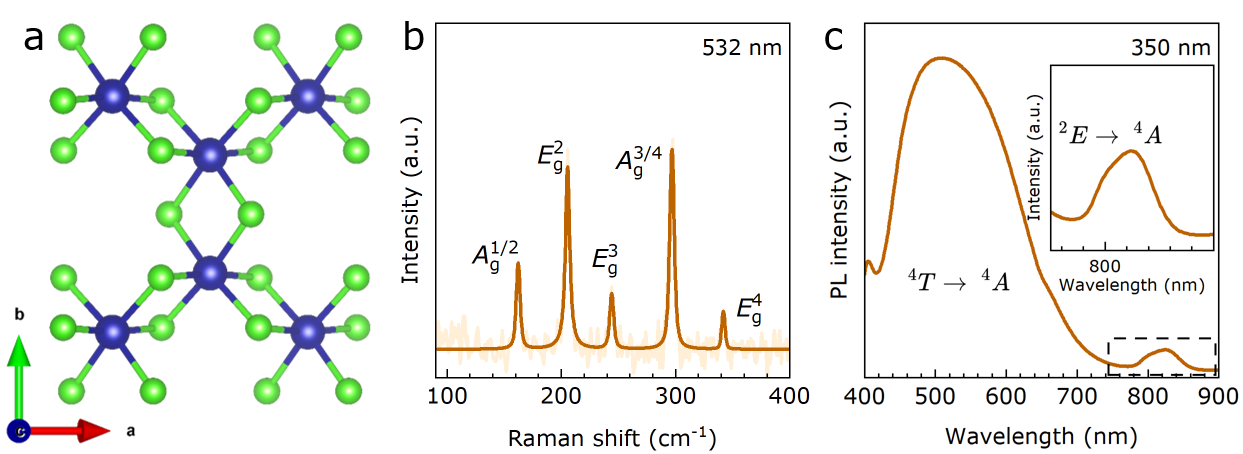}
    \caption{Optical characterization of bulk CrCl$_3$. (a) Schematic unit cell of monoclinic CrCl$_3$. Here, blue and green balls represent Cr and Cl atoms, respectively. (b) Raman spectrum recorded under 532 nm excitation, showing the prominent Raman-active modes. The experimental data are shown as the light shaded curve, while the solid line represents a Voigt fit of the spectrum. Here, the $A_\textrm{g}^{1,2}$ and $A_\textrm{g}^{3,4}$ are the nearly degenerate phonon modes. (c) PL of CrCl$_3$ measured using 350 nm excitation. The broad emission centered in the Vis range is attributed to the spin-allowed $^4T \rightarrow {}^4A$ transition. The weak NIR feature highlighted by the dashed box is shown enlarged in the inset and corresponds to the spin-forbidden $^2E \rightarrow {}^4A$ transition.}
    \label{fig:Raman_PL}
\end{figure}

Figure \ref{fig:Raman_PL}b displays the Raman spectrum of a representative CrCl$_3$ flake measured under ambient conditions. Five prominent Raman modes are present, which can be assigned to $A_\textrm{g}$ ($A_\textrm{g}^1$ - $A_\textrm{g}^4$) and $E_\textrm{g}$ ($E_\textrm{g}^2$ - $E_\textrm{g}^4$) vibrations of the layered lattice, consistent with a recent report on CrCl$_3$ \cite{Kipczak2026}. The $A_\textrm{g}$ modes correspond to fully symmetric lattice vibrations involving predominantly out-of-plane atomic displacements, whereas the $E_\textrm{g}$ modes arise from in-plane vibrations transforming according to the two-dimensional representation of the layered crystal symmetry. Notably, the $A_\textrm{g}^{1/2}$ and $A_\textrm{g}^{3/4}$ features appear as two closely spaced doublets, consistent with pairs of nearly degenerate $A_\textrm{g}$ phonons reported for multilayer CrCl$_3$ \cite{Kipczak2026}. The sharp linewidths and well-defined peak positions indicate good crystalline quality and minimal structural degradation, confirming that the intrinsic lattice symmetry is preserved.\par
The PL spectrum shown in Figure \ref{fig:Raman_PL}c reveals two emission features in the Vis and NIR regions. The higher-energy feature at about 510 nm originates from the $^4T \rightarrow {}^4A$ transition \cite{Polini1970, Li2025_2}. The lower-energy emission near 830 nm is attributed to the spin-forbidden $^2E \rightarrow {}^4A$ (d-d) transition, weakly activated by spin-orbit coupling and lattice distortions, and characteristic of Cr$^{3+}$-based octahedral systems \cite{Polini1970, Cai2019, Li2025_2, Seyler2018}.

\subsection{High-performance photodetection enabled by long-lived excitons}\label{result_power}
\begin{figure}[h]
    \centering
    \includegraphics[width=1.0\textwidth]{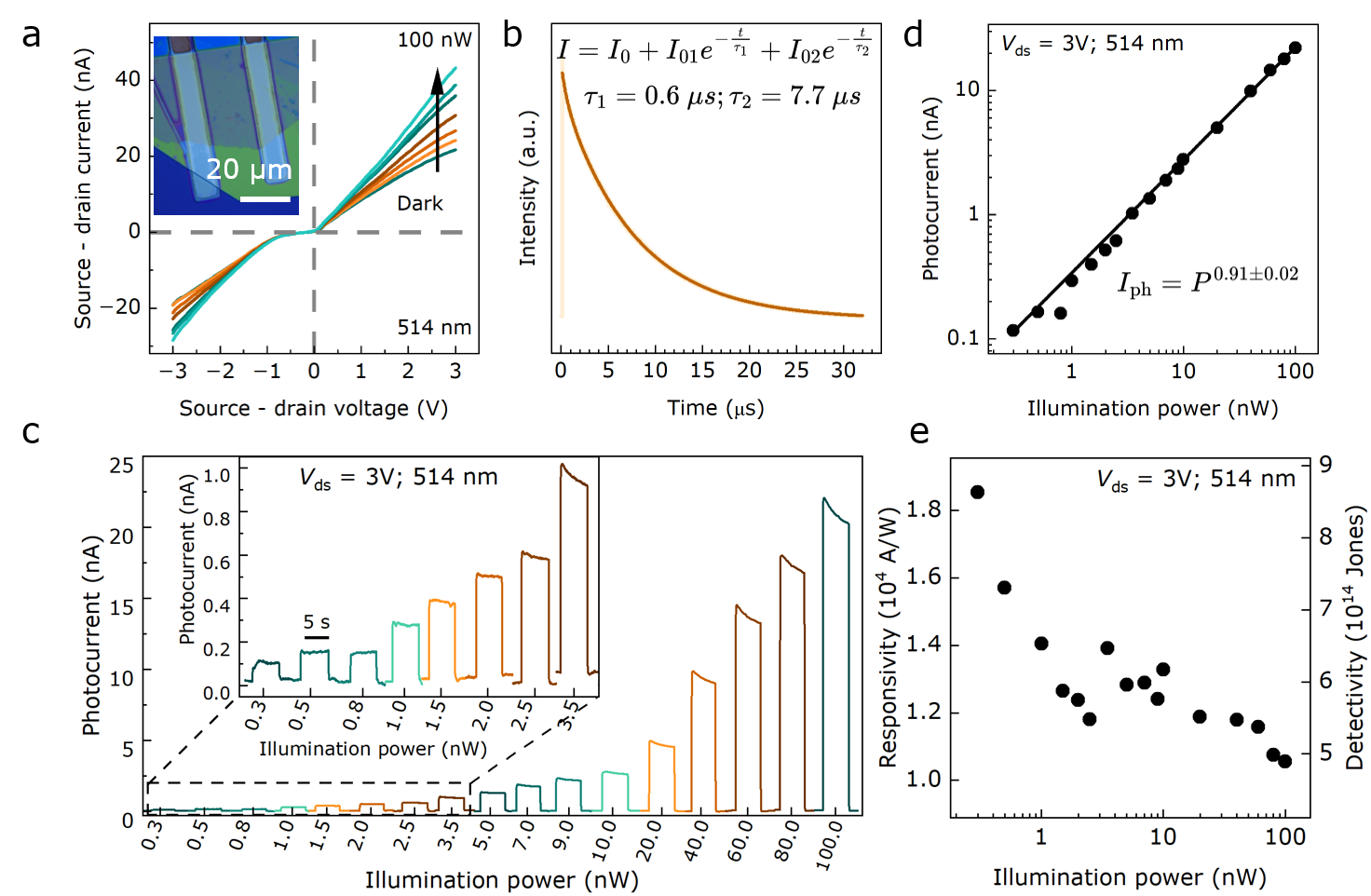}
    \caption{Optoelectronic response of bulk CrCl$_3$ flake under 514 nm illumination. (a) Source-drain current--voltage ($I_\textrm{ds}-V_\textrm{ds}$) characteristics measured in the dark and under illumination at different excitation powers (upto 100 nW), showing nearly symmetric transport and pronounced photocurrent enhancement without rectification. Inset: optical image of the device. Scale bar: 20 {\textmu}m. (b) TRPL of CrCl$_3$ under 532 nm illumination fitted using the biexponential function ($I = I_0 + I_{01}e^{-t/\tau_1} + I_{02}e^{-t/\tau_2}$) showing the large exciton recombination lifetime of  $\tau_2 = 7.7 ~\mu$s for $^2E \rightarrow {}^4A$ transition. (c) Time-resolved photocurrent under periodic monochromatic illumination at different excitation powers. The inset highlights a photocurrent signal under a sub-nW power range. (d) Photocurrent as a function of laser power fitted using $I_\textrm{ph} \propto P^{\alpha} (\alpha = 0.91)$. (e) Responsivity and detectivity as a function of laser power.}
    \label{fig:high_performance}
\end{figure}
Figure \ref{fig:high_performance} summarizes the high-performance photodetection characteristics of the CrCl$_3$ device under 514 nm illumination. The device was fabricated using deterministic PDMS-assisted dry transfer of mechanically exfoliated flakes onto pre-patterned Cr/Au electrodes (see Experimental Section for details).\par
Figure \ref{fig:high_performance}a shows the source-drain current ($I_\textrm{ds}$) as a function of the applied bias ($V_\textrm{ds}$) measured in the dark and under 514 nm illumination at different optical powers in the range of 0.3 nW to 100 nW (0.4 nW/cm$^2$ to 141 nW/cm$^2$). In the dark, the $I_\textrm{ds}$ remains as low as $\sim$ 20 nA at 3 V, reflecting the highly resistive nature of the material and the absence of significant free-carrier under equilibrium conditions \cite{Cai2019}. Upon illumination, however, the $I_\textrm{ds}$ increases rapidly and reproducibly. The $I_\textrm{ds}-V_\textrm{ds}$ curves remain nonlinear under illumination, indicating that the photocurrent does not arise from a simple ohmic photoconductive response but instead reflects field-assisted extraction of photoexcited carriers. Importantly, the large photocurrent contrast between dark and illuminated conditions demonstrates that photon illumination dramatically alters the effective carrier population available for transport.\par
The origin of this pronounced photoresponse becomes clear when viewed together with the time-resolved photoluminescence (TRPL) data shown in Figure \ref{fig:high_performance}b. The TRPL decay can be well described by a biexponential function, yielding a dominant long lifetime of $\tau_2 = 7.7 ~\mu$s, along with a faster initial component ($\tau_1 = 0.6 ~\mu$s) \cite{Sridhar2024}. In CrCl$_3$, the faster component is likely linked to the initial high-density excitonic population, where exciton-exciton annihilation contributes significantly to the decay, while the slower component reflects the intrinsic single-exciton recombination lifetime \cite{Sridhar2024, Snoeren2023}. This extended recombination lifetime allows photoexcited carriers to survive long enough to drift under the applied electric field and be collected at the electrodes before recombination, resulting in efficient carrier extraction despite the intrinsically low mobility ($< 0.05$ cm$^2$/Vs) of CrCl$_3$ due to its insulating nature \cite{Cai2019, Lin2019}.\par
Figure \ref{fig:high_performance}c presents the power-dependent photocurrent measured at $V_\textrm{ds} = 3$ V under 514 nm illumination, spanning nearly three orders of magnitude in optical power. Clear and stable ON--OFF switching is observed down to an illumination power of 0.3 nW (SNR $\approx$ 4), where a distinct, low-noise photocurrent remains readily detectable during a 5-second interval. The photocurrent increases monotonically with optical power (Figure \ref{fig:high_performance}d), reflecting efficient photogeneration and extraction across the entire measured range. A power law fit yields $I_\textrm{ph} \propto P^{0.91\pm0.02}$, indicating a nearly linear response, which confirms that the device operates far from saturation over the measured power range. The slight sublinearity suggests weak recombination or trap-assisted effects at higher carrier densities \cite{Haider2016}.\par
The sensitivity of the device is quantified through the standard photodetector figures of merit, shown in Figure \ref{fig:high_performance}e (see Supporting Information for details). At low illumination powers, the responsivity exceeds $1.8 \times 10^4$ A/W, corresponding to an internal photocurrent gain of approximately $4.5 \times 10^4$. The exceptionally high responsivity should not be interpreted as evidence of high intrinsic carrier mobility. Instead, it arises from the interplay between the long-lived ligand-field excitons and the applied electric field. The microsecond-scale excited-state lifetime provides an extended temporal window for exciton dissociation, field-assisted carrier extraction, and trap-mediated photogating. Long-lived trapped or metastable populations can maintain an enhanced channel conductivity during photoexcitation, thereby producing large photoconductive gain despite the low intrinsic mobility. Moreover, this exciton lifetime could substantially exceed the carrier transit time across the device channel, and the photoexcited carriers can undergo multiple drift cycles before recombination, giving rise to large photoconductive gain \cite{Haider2016}. This gain mechanism distinguishes CrCl$_3$ from conventional semiconductor photodetectors, where high responsivity is typically achieved through efficient carrier transport rather than prolonged carrier lifetime. The responsivity gradually decreases with increasing optical power, consistent with the sublinear photocurrent scaling observed in Figure \ref{fig:high_performance}d. This further indicates trap filling and recombination-limited transport at higher carrier density, suggesting the presence of gain mechanisms beyond simple photoconduction \cite{Haider2016}.\par

Using the measured dark current, the specific detectivity reaches a maximum value of $8.6 \times 10^{14}$ Jones at low illumination powers under the assumption of shot-noise limited operation (see Supporting Information for details). This notably high detectivity reflects the combined effect of a large photocurrent gain and low-noise ($I_\textrm{dark} = 21.7$ nA at $V_\textrm{ds} = 3$ V) current. With increasing illumination power, the detectivity shows a moderate reduction, following the decrease in responsivity, while remaining above $4.8 \times 10^{14}$ Jones over a broad power range. These values place CrCl$_3$ among the highly sensitive photodetectors reported for layered systems \cite{Han2025, Abdullah2024}, and notably competitive with state-of-the-art photodetectors based on a similar class of materials (also see Supporting Table 1), making CrCl$_3$ a promising platform for future optoelectronic applications, particularly in the low-power regime relevant to weak-light and low-flux detection conditions.

\subsection{Ultrabroadband optical response from UV to NIR (300--1700 nm)}\label{result_wavelength}
\begin{figure}[h]
    \centering
    \includegraphics[width=1.0\textwidth]{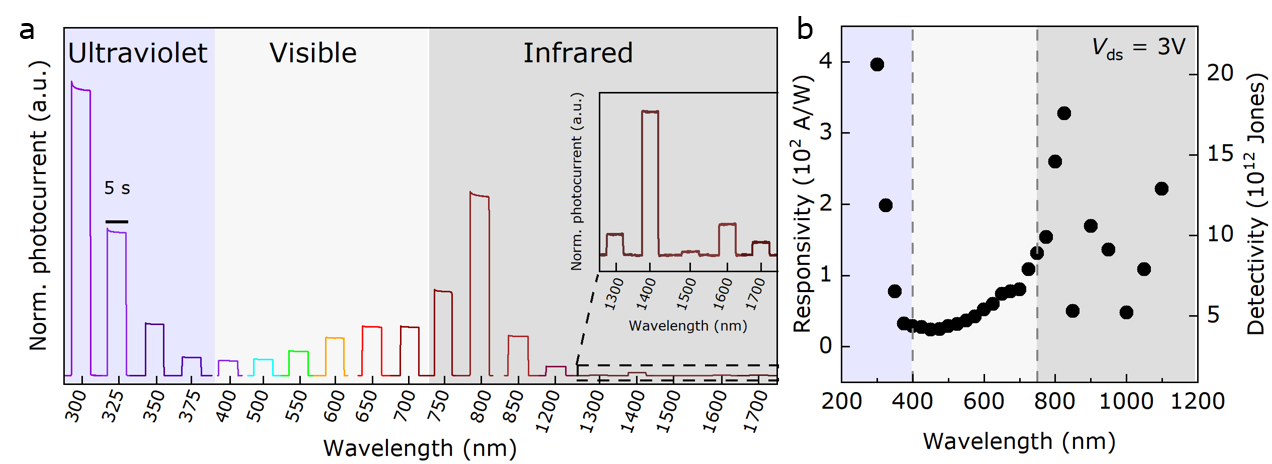}
    \caption{Ultrabroadband optoelectronic response of CrCl$_3$. (a) Time-resolved photocurrent (normalized w.r.t. to illumination power) under periodic monochromatic illumination across UV, Vis, and NIR wavelengths. The inset highlights a photocurrent signal extending into the 1300-1700 nm range. (b) Responsivity and detectivity throughout the spectral region.}
    \label{fig:broadband}
\end{figure}
Figure \ref{fig:broadband} explores the wavelength-dependent photoresponse of CrCl$_3$ over a broad spectral range, from the ultraviolet (UV) to the NIR, and links the observed photocurrent to the underlying excitonic and emissive processes.\par
The wavelength-resolved temporal photoresponse is shown in Figure \ref{fig:broadband}a, where the device is periodically illuminated with monochromatic light spanning the UV (300--400 nm), Vis (400--750 nm), and NIR (750--1700 nm) regions. Robust and repeatable photocurrent signals are observed across the entire spectral range, with clear ON--OFF switching under periodic illumination. While the absolute photocurrent magnitude varies with wavelength, the persistence of a measurable signal even at long wavelengths (up to 1700 nm) demonstrates genuine ultrabroadband sensitivity. At 1700 nm, the photocurrent remains above the noise floor with a signal-to-noise ratio of $\sim$10, confirming that the response is well within the detection limit. This broadband response indicates that photocurrent generation in CrCl$_3$ does not rely on a single interband transition, but instead involves a hierarchy of excitonic and weakly bound states that can be accessed over a wide excitation-energy window \cite{Polini1970, Zhu2020, Abdel2025}.\par
The corresponding responsivity and detectivity are shown in Figure \ref{fig:broadband}b. Both quantities exhibit pronounced wavelength dependence, reflecting variations in optical absorption, exciton formation, and carrier extraction efficiency across the spectrum \cite{Polini1970, Zhu2020, Abdel2025}. The responsivity reaches its maximum in the UV region around 300 nm, where excitation occurs well above the fundamental absorption edge. In this spectral range, strong interband absorption generates relatively delocalized charge carriers with high mobility, enabling efficient charge separation and collection and resulting in a large photocurrent gain, in agreement with a previous report \cite{Polini1970}.\par
In the Vis region, the responsivity is reduced to approximately one quarter of its maximum value. This reduction can be attributed to the nature of the optical absorption in this energy range \cite{Polini1970}, where enhanced absorption leads to carrier generation predominantly near the surface. As a result, the effective carrier transport length is reduced, limiting the contribution of these carriers to the measured photocurrent \cite{Polini1970}. In the NIR region, the photoresponse increases again, consistent with excitation involving ${}^4A \rightarrow ^2E$ electronic states, before exhibiting oscillatory features at longer wavelengths due to multiphonon absorption, in line with reported phonon-assisted absorption features in insulating crystals \cite{Sparks1973}. These features indicate a sub-bandgap photoresponse dominated by low-efficiency excitation and transport channels, in agreement with earlier absorption and photoconductivity studies \cite{Polini1970, Kipczak2026}.\par
It is also important to note that the responsivity values shown here are obtained under moderate illumination powers (20 {\textmu}W - 700 {\textmu}W) used for wavelength-dependent measurements. The responsivity values in this case are, therefore,  almost two orders of magnitude lower than what was discussed earlier (in section \ref{fig:high_performance}), reflecting the transition to a higher-power steady-state regime.

\subsection{In-plane polarization selectivity}\label{result_in_plane}
\begin{figure}[htbp!]
    \centering
    \includegraphics[width=1\linewidth]{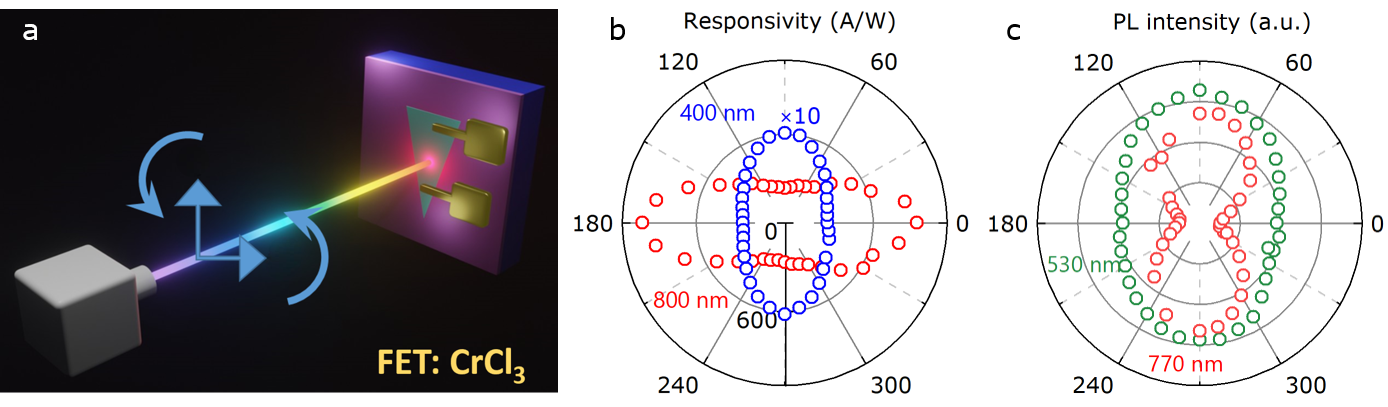}
    \caption{In-plane anisotropy of photoresponse and emission in CrCl$_3$. (a) Schematic of the polarization-resolved measurement geometry. The linear polarization of the excitation beam was rotated with respect to the crystal axes while monitoring either photocurrent or emission. (b) Polarization dependence of responsivity under 400 nm (blue) and 800 nm (red) excitation. (c) Polarization dependence of the 828 nm PL under 530 nm at 500 {\textmu}W (green) and 770 nm at 100 {\textmu}W (red) excitation.}
    \label{fig:in-plane}
\end{figure}
The strong wavelength dependence suggests that multiple optical transitions contribute to the photocurrent. To identify their polarization characteristics, we next investigate the in-plane anisotropy in CrCl$_3$ photodetectors. The layered crystal possesses low in-plane symmetry, and this structural anisotropy is expected to leave a fingerprint on the optical absorption and photocurrent generation. Figure \ref{fig:in-plane}a shows the schematic of the experimental configuration, in which the linear polarization of the incident light was rotated, and the photocurrent and PL were measured.\par
Figure \ref{fig:in-plane}b shows the polarization dependence of the responsivity under 400 nm and 800 nm excitation. In both cases, the response exhibits a pronounced twofold symmetry, consistent with the monoclinic structure of CrCl$_3$. Excitation at 400 nm produces a relatively low, yet strongly polarization-modulated, response, with maxima at about 90\textdegree. In contrast, 800 nm excitation yields nearly an order-of-magnitude higher responsivity, with polarization maxima rotated by roughly 90\textdegree~relative to those at 400 nm. This shift suggests that different optical transitions dominate the photoresponse at the two excitation energies: higher-energy states at 400 nm couple preferentially to one axis, while near-resonant ligand-field states at 800 nm couple along the orthogonal axis, enhancing carrier generation efficiency \cite{Ho2019}. The observed orientation dependence is consistent with the in-plane dipole anisotropy of the transitions. \par
The orthogonal polarization axes observed for the 400 nm and 800 nm photoresponses provide important insight into the underlying optical transitions. In an ideal CrCl$_3$ crystal, the in-plane optical response is expected to be nearly isotropic owing to the threefold symmetry of the honeycomb lattice, with the primary optical anisotropy occurring between the in-plane ($ab$) and out-of-plane ($c$) directions. The pronounced polarization dependence observed here, therefore, suggests that the effective optical response is governed by lower-symmetry dipole channels than would be expected for an ideal crystal. Such behavior indicates that specific optical transitions possess preferential dipole orientations, giving rise to distinct polarization-selective excitation pathways. Within this framework, the lower-energy optical response near 800 nm is associated with localized ligand-field (Cr$^{3+}$ d-d) excitons, whereas higher-energy excitations in the Vis range are attributed to charge-transfer or higher-lying crystal-field states involving Cl p--Cr d hybridization. Because these transitions possess different orbital character, their transition dipole moments can preferentially align along different principal axes of the distorted lattice, naturally accounting for the nearly orthogonal polarization maxima observed in the photocurrent measurements. Such behavior is consistent with previous theoretical and experimental studies highlighting the strong excitonic character of chromium trihalides and the sensitivity of their optical selection rules to symmetry breaking \cite{Polini1970, Zhu2020}.\par
To probe the radiative recombination channels, the polarization dependence of the emission at 828 nm was measured under two excitation energies, 530 nm and 770 nm (Figure \ref{fig:in-plane}c). In both cases, the emission exhibits a clear twofold modulation, aligned with the polarization axis observed for the 400 nm responsivity. Importantly, the overall shape and orientation of the polarization pattern remain nearly similar for the two excitation wavelengths, indicating that relaxation processes funnel carriers into the same emissive state whose dipole orientation is fixed by the local ligand-field symmetry. While the angular dependence is preserved, the degree of anisotropy changes, with a broader polarization profile and enhanced intensity along the principal emission axis under 530 nm excitation compared to 770 nm excitation. This excitation-energy-dependent modulation of the anisotropy amplitude is attributed to resonance effects that alter the efficiency of carrier generation and relaxation pathways without changing the symmetry of the terminal emissive state. Thus, while the photocurrent anisotropy is strongly dependent on the excitation energy and reflects the dipole orientation of the initial absorption channel, the emission anisotropy remains locked to the symmetry of the terminal emissive $^2$E state. \par
These in-plane anisotropy measurements reveal that the photocurrent is intimately tied to the orientation of excitonic dipoles. The results highlight the importance of exciton-mediated processes in determining both magnitude and polarization dependence of the photoresponse. They also establish a baseline for understanding three-dimensional dipole contributions, which are explored through sample angle-dependent out-of-plane measurements in the next section. 
\subsection{Out-of-plane polarization sensitivity}\label{result_out_of_plane}
\begin{figure}[htbp!]
    \centering
    \includegraphics[width=1\linewidth]{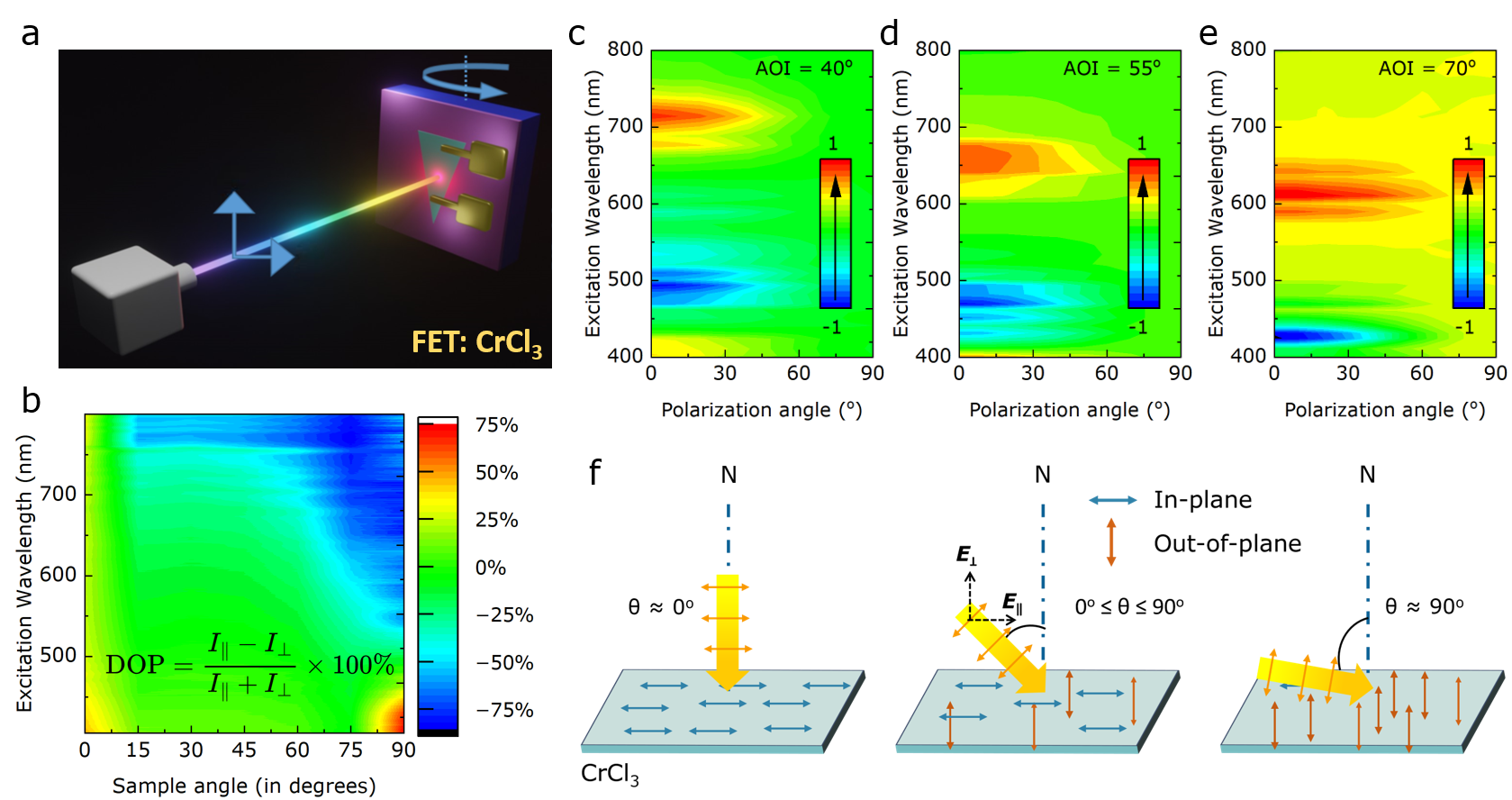}
    \caption{Out-of-plane anisotropy of the photocurrent response. (a) Schematic of the experimental geometry used to probe out-of-plane anisotropy. The angle of incidence (AOI) of the excitation beam was varied with respect to the sample, introducing a controllable out-of-plane electric field component while maintaining fixed incident polarization. (b) Degree of polarization ($DOP = \frac{I_\parallel - I_\perp}{I_\parallel + I_\perp} \times 100 \%$) of the photocurrent, plotted as a function of excitation wavelength and AOI. (c-e) Reflectivity as a function of polarization angle and excitation wavelength for different AOI values. (f) Schematic of the working mechanism.}
    \label{fig:out-of-plane}
\end{figure}
To understand the role of out-of-plane optical dipoles in CrCl$_3$, we performed angle-resolved polarization-dependent photocurrent and reflection measurements using the geometry illustrated in Figure \ref{fig:out-of-plane}a. In this configuration, the angle of incidence (AOI) with respect to the surface normal is varied, thereby tuning the relative projection of the incident electric field onto the in-plane (E$_\parallel$) and out-of-plane (E$_\perp$) components. This approach allows us to disentangle the contributions of in-plane and out-of-plane transition dipoles to light absorption and photocurrent generation.\par
Figure \ref{fig:out-of-plane}b presents the photocurrent degree of polarization (DOP) as a function of excitation wavelength and sample rotation angle, calculated from the $I_{\parallel}$ and $I_{\perp}$ photocurrent components shown in the Supporting Figure S1. The DOP shows a pronounced dependence on both parameters, with values spanning approximately from $-$90\% to $+$75\% across the measured spectral range. At small AOI $\leq$ 10\textdegree, the DOP remains close to zero over most wavelengths, indicating a weak polarization selectivity of the photocurrent under near-normal incidence. As the sample angle increases, the DOP magnitude progressively grows, reaching large positive or negative values for AOI $\geq$ 60\textdegree, depending on the excitation wavelength.\par
Distinct spectral regions with opposite DOP sign are observed as a function of angle. In the UV range, the DOP tends toward positive values at large angles, whereas in the Vis, a sign reversal occurs and the DOP becomes strongly negative. The systematic enhancement of \textbar DOP\textbar~with increasing sample angle suggests that the photocurrent response cannot be described by a purely in-plane optical dipole model. Instead, the data reveal a substantial contribution from out-of-plane optical dipoles, whose coupling strength increases as the incident electric field acquires a larger out-of-plane component E$_\perp$ at oblique incidence.\par
This interpretation is corroborated by the reflection intensity maps (details in the Experimental Section) shown in Figure \ref{fig:out-of-plane}c-e, measured at AOI = 40\textdegree, 55\textdegree, and 70\textdegree, respectively (see Supporting Figure S2 for the AOI = 45\textdegree, 50\textdegree, 60\textdegree, and 65\textdegree). At a smaller AOI, the reflection anisotropy is dominated by features associated with in-plane dipoles (Figures \ref{fig:out-of-plane}c and \ref{fig:out-of-plane}f, left panel), producing polarization-dependent intensity maxima primarily in the Vis range. As the AOI increases, the features shift both in polarization angle and excitation wavelength, reflecting the growing contribution of out-of-plane dipole transitions (Figures \ref{fig:out-of-plane}d-e and \ref{fig:out-of-plane}f, middle and right panels). At the largest AOI = 70\textdegree, the reflection response becomes strongly polarized even in spectral regions that are weakly anisotropic at lower angles, providing clear optical evidence for transitions with a substantial out-of-plane dipole moment.\par
Within a linear dipole-superposition framework, the absorbed power, and hence the photocurrent, can be expressed as
\begin{equation}
    I_\textrm{ph} (\lambda, \theta) \propto |\mu_{\parallel}|^2 |E_{\parallel} (\theta)|^2 + |\mu_{\perp}|^2 |E_{\perp} (\theta)|^2
\end{equation}
where $\mu_{\parallel}$ and $\mu_{\perp}$ are the in-plane and out-of-plane transition dipole moments, respectively, and $\theta$ denotes the AOI. For the sake of simplicity, interference, multiple reflections, and cross terms are neglected in this qualitative phenomenological model, and more sophisticated modelling (including that of the anisotropic dielectric tensor and Fresnel coefficients) is beyond the scope of this work.\par
This framework rationalizes both the enhancement of the photocurrent DOP and its pronounced evolution with angle of incidence. As the AOI increases, the out-of-plane electric field component $E_{\perp}$ grows relative to the in-plane component $E_{\parallel}$, thereby selectively amplifying transitions with finite out-of-plane dipole moments (Figure \ref{fig:out-of-plane}f). Consequently, optical channels that are weak or inactive under near-normal incidence become increasingly dominant, leading to a redistribution of the photocurrent among competing dipole contributions. This progressive reweighting enhances the polarization contrast because the in-plane and out-of-plane dipoles generally exhibit different angular and spectral dependencies. Moreover, since these dipole channels are associated with distinct electronic transitions, their relative contributions vary with excitation energy. This results not only in a monotonic increase in the magnitude of the DOP but also in its sign reversal across different spectral regions, reflecting a crossover between regimes where either in-plane or out-of-plane dipoles dominate the photoresponse. The combined effect is consistent with a strong, angle-tunable polarization sensitivity that encodes the vectorial nature of the underlying optical transitions. \par
Taken together, the wavelength-, polarization-, and angle-resolved measurements establish a unified picture in which the optoelectronic response of CrCl$_3$ is governed by an energy-dependent three-dimensional excitonic dipole landscape. Rather than representing isolated optical transitions, the observed polarization behavior reflects the continuous redistribution of excitonic oscillator strength among competing in-plane and out-of-plane dipole channels. This framework naturally accounts for the ultrabroadband photoresponse, excitation-dependent polarization rotation, and giant tunability of the degree of polarization, providing a unified physical description of vectorial light-matter interactions in layered magnetic insulators.
\section{Conclusions}\label{conclusion}
In summary, we demonstrate that the intrinsic dielectric anisotropy of layered CrCl$_3$ gives rise to a three-dimensional excitonic dipole landscape that simultaneously enables ultrabroadband photodetection and intrinsic polarization selectivity. This unique excitonic landscape underpins high-performance photodetection spanning 300 to 1700 nm, exhibiting a responsivity exceeding $1.8 \times 10^4$ A/W with a photoconductive gain above $4.5 \times 10^4$. By combining wavelength-, polarization-, and angle-resolved optoelectronic measurements with polarization-resolved spectroscopy, we reveal that distinct ligand-field and higher-energy excitonic transitions possess different optical dipole orientations, leading to excitation-energy-dependent rotation of the polarization axis and the activation of out-of-plane optical dipoles under oblique illumination. Together, these observations reveal that the polarization response is governed by the energy-dependent redistribution of excitonic oscillator strength between in-plane and out-of-plane dipole channels. These findings establish a direct relationship between excitonic dipole orientation and vectorial photocurrent generation, demonstrating that optical anisotropy in magnetic van der Waals materials extends beyond conventional in-plane selection rules. \par
More broadly, our results demonstrate that intrinsic dielectric anisotropy provides a powerful route for engineering vectorial light-matter interactions without artificial nanostructuring or metasurfaces. Extending this concept to magnetic van der Waals heterostructures, moiré superlattices, and electrically tunable architectures could enable programmable polarization-selective optical functionality. Beyond photodetection, these findings position excitonic dipole engineering as a general design strategy for integrated photonic and quantum optical technologies. 

\section{Experimental section}\label{methods}
\textit{CrCl$_3$ crystal synthesis}: CrCl$_3$ crystals were synthesized on SiO$_2$ (300 nm)/Si substrates via a chemical vapor transport growth method. Anhydrous CrCl$_3$ powder (99.99\%, Sigma-Aldrich) was loaded into a quartz boat, and a SiO$_2$/Si substrate was placed face-down above the precursor material. The assembly was inserted into a quartz tube furnace and heated to 820 \textdegree C at a ramp rate of 40 \textdegree C/min under a continuous argon (95\%) flow of 100 sccm. After holding the peak temperature for one minute, the furnace was switched off and allowed to cool naturally to room temperature. Following cooling, the substrates bearing plate-like crystalline flakes were collected for further characterization.\\
\textit{Exfoliation and device fabrication}: Single-crystalline CrCl$_3$ photodetectors were prepared using a dry-transfer approach on heavily p-doped SiO$_2$ (300 nm)/Si substrates. Electrical contacts were defined on the SiO$_2$/Si substrate via maskless laser lithography (MicroWriter, Durham Magneto Optics Ltd.), followed by metal deposition and lift-off. The metal layers consisting of 5 nm Cr and 45 nm Au were deposited by sputtering to form the electrodes. The excess metal was removed by lift-off in heated acetone and subsequently cleaned with isopropyl alcohol. Separately, CrCl$_3$ flakes were mechanically exfoliated onto a thin PDMS (Gel-Pak 4) film using a low-adhesion tape (Nitto Denko Corp.). Selected flakes were then deterministically transferred from the PDMS stamp onto the pre-patterned substrate using a micromanipulator-assisted transfer stage \cite{Panda2023}. The completed devices were examined under an optical microscope, with imaging conditions optimized to enhance the optical contrast between CrCl$_3$ and the SiO$_2$/Si substrate.\\
\textit{Raman and photoluminescence measurements}: Raman measurements were performed using a confocal micro-Raman setup (WITEC Alpha 300R) equipped with a 1800 lines/mm grating in a backscattering geometry. A continuous-wave laser excitation at 2.33 eV (532 nm) was used as the excitation source. The laser was focused onto the sample using a high numerical aperture (NA = 0.9) objective, resulting in a diffraction-limited spot size of approximately 0.5 {\textmu}m. To minimize local heating and photoinduced effects, the incident laser power was kept at 100 {\textmu}W. PL measurements were carried out using a Fluorolog spectrofluorometer (Horiba Jobin-Yvon Fluorolog-3) equipped with a 1200 lines/mm grating. A mercury vapor lamp was used as the excitation source at 350 nm.\\
\textit{Time-resolved photoluminescence measurements}: Time-resolved photoluminescence (TRPL) measurements were carried out at room temperature using a 532 nm pulsed laser with a pulse width of 20 ps and a repetition rate of 1 MHz. The emitted PL was collected in a polarization-resolved confocal microscope geometry and analyzed using a time-correlated single-photon counting setup equipped with two single-photon detectors. The overall instrument response function of the system was 125 ps. To avoid detector saturation and nonlinear effects arising from the strong emission of CrCl$_3$, the excitation power was kept below 100 nW, as calibrated using a continuous-wave laser power meter. Measurements were performed on a polarization-resolved confocal microscope platform operated under vacuum ($\sim~10^{-6}$ torr).\\
\textit{Electrical and photodetection measurement}: Electrical transport and photodetection measurements were carried out using a Fluorolog spectrofluorometer (Horiba Jobin-Yvon 3) as a broadband, wavelength-tunable light source. Monochromatic excitation was selected from the mercury vapor lamp (300-1100 nm) using the excitation monochromator, and the resulting photocurrent was recorded as a function of excitation wavelength under an applied bias. The electrical biasing and current readout were performed using a Keithley 2610 sourcemeter. A large illumination spot size of approximately 5 mm was used to ensure uniform excitation over the device area and to minimize spatial inhomogeneity effects.\\
\textit{Reflectivity measurements}: Reflectivity measurements were performed using the EP4 imaging ellipsometer (Park Systems). The illumination source was an Xe lamp with a filter wheel for the selection of a narrow illumination bandwidth (approximately 10 nm), providing 45 wavelengths in the range from 360 to 1000 nm. The signal was acquired with a CCD camera, which allows signal averaging from certain sample regions. Polarization of the light was controlled with the position of the linear polarization filter and quarter-wave plate (aligned to not distort the linear polarization) in the illumination branch. The detector branch was equipped with a 10$\times$ objective for better spatial resolution and another linear polarization filter (in front of the CCD camera), set to the same polarization plane as the optical elements in the illumination branch. The positions of the optical elements, the setting of the illumination wavelength, and the acquisition of signals were controlled with an external Python script. The angle of incidence was manually set from 40\textdegree ~to 70\textdegree ~with an increase of 5\textdegree.\\
All the experiments were carried out at a room temperature under ambient conditions.

\backmatter
\bmhead{Acknowledgements}
S.S. and M.V. acknowledge the support of the Lumina Quaeruntur fellowship No. LQ200402201 by the Czech Academy of Sciences. G.H. acknowledges financial support through start-up funding from Leibniz IFW Dresden. This work was also supported by the Ministry of Education, Youth, and Sports of the Czech Republic, Project No. CZ.02.01.01/00/22\_008/0004558, co-funded by the European Union, and the CzechNanoLab Research Infrastructure, supported by the Ministry of Education, Youth, and Sports of the Czech Republic (LM2023051).\\

\textbf{Data availability:} All the data generated and analyzed in this study are included in the Article and its Supplementary Information. The data that supports the plots within this paper and other findings of this study are available from the Catch-all at [persistent link to data repository will be provided upon acceptance]. 


\printbibliography
\newpage
\subsubsection*{Table of contents graphic}
\flushleft{Layered CrCl$_3$ enables ultrabroadband photodetection from ultraviolet to near-infrared wavelengths while simultaneously resolving light polarization without external nanostructuring. The response originates from dielectric anisotropy and distinct optical dipoles that govern light--matter interaction across different energies. The device exhibits strongly tunable polarization sensitivity, demonstrating the potential of van der Waals insulator for compact multifunctional optoelectronic and photonic technologies.}
\begin{figure}[h]
    \centering
    \includegraphics[width=5.5cm]{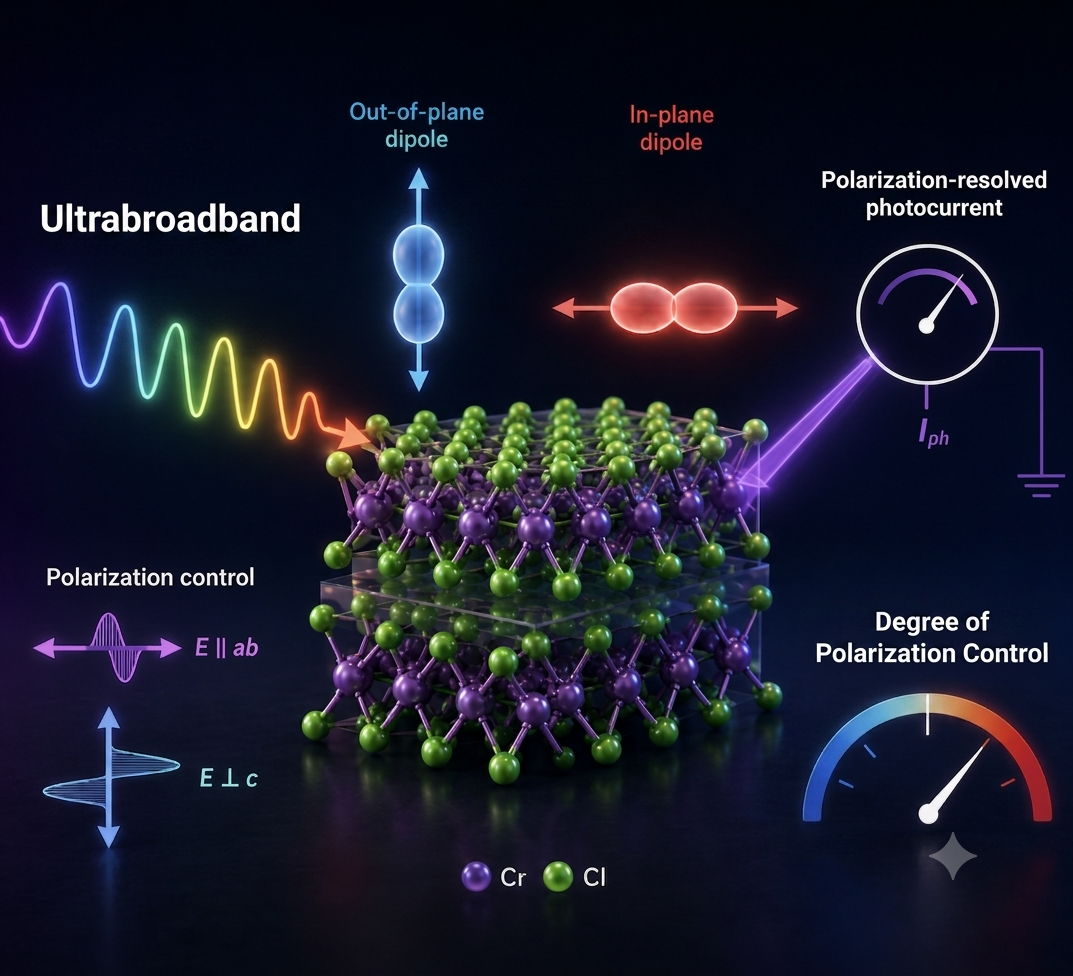}
\end{figure}


\end{document}


\title[Article Title]{Supporting Information: Three-dimensional excitonic dipole anisotropy enables ultrabroadband polarization photodetection in CrCl$_3$}

\author[1]{\fnm{Satyam} \sur{Sahu}}\email{satyam.sahu@jh-inst.cas.cz}
\author[1,2]{\fnm{Jaganandha} \sur{Panda}}\email{jagan.panda@uni-muenster.de}
\author[1]{\fnm{Martin} \sur{Jindra}}\email{martin.jindra@jh-inst.cas.cz}
\author[1,3]{\fnm{Mukesh} \sur{Kumar Thakur}}\email{mukesh@iiitvadodara.ac.in}
\author[1,4]{\fnm{Farjana J.} \sur{Sonia}}\email{f.j.sonia@ifw-dresden.de}
\author[5]{\fnm{Shankar} \sur{Khanal}}\email{skhanal@mag.mff.cuni.cz}
\author[4]{\fnm{Kornelius} \sur{Nielsch}}\email{k.nielsch@ifw-dresden.de}
\author[5]{\fnm{Jana} \sur{Vejpravova}}\email{jana.vejpravova@matfyz.cuni.cz}
\author[1]{\fnm{Matěj} \sur{Velický}}\email{matej.velicky@jh-inst.cas.cz}
\author*[1]{\fnm{Martin} \sur{Kalbáč}}\email{martin.kalbac@jh-inst.cas.cz}
\author*[1]{\fnm{Otakar} \sur{Frank}}\email{otakar.frank@jh-inst.cas.cz}
\author*[1,4]{\fnm{Golam} \sur{Haider}}\email{g.haider@ifw-dresden.de}

\affil*[1]{\orgname{Heyrovský Institute of the Czech Academy of Sciences}, \orgaddress{\street{Dolejškova 2155/3}, \city{Prague}, \postcode{18200}, \country{Czech Republic}}}

\affil[2]{\orgname{Münster Nanofabrication Facility, University of Münster}, \orgaddress{\street{Busso-Peus-Str.10}, \city{Münster}, \postcode{48149}, \country{Germany}}}


\affil[3]{\orgname{Department of Applied Physics, Indian Institute of Information Technology Vadodara}, \orgaddress{\city{Gandhinagar}, \postcode{382028}, \country{India}}}

\affil*[4]{\orgname{Leibniz Institute for Solid State and Materials Research}, \orgaddress{\street{Helmholtzstr. 20}, \city{Dresden}, \postcode{01069}, \country{Germany}}}

\affil[5]{\orgname{Department of Condensed Matter Physics, Charles University}, \orgaddress{\street{Ke Karlovu 5}, \city{Prague}, \postcode{12116}, \country{Czech Republic}}}


\maketitle

\pagebreak
\clearpage
\renewcommand{\thefigure}{S\arabic{figure}}
\setcounter{figure}{0}
\renewcommand{\thepage}{S-\arabic{page}}
\setcounter{page}{1}
\renewcommand{\theequation}{S.\arabic{equation}}
\setcounter{equation}{0}

\begin{figure}[htbp!]
    \centering
    \includegraphics[width=0.8\linewidth]{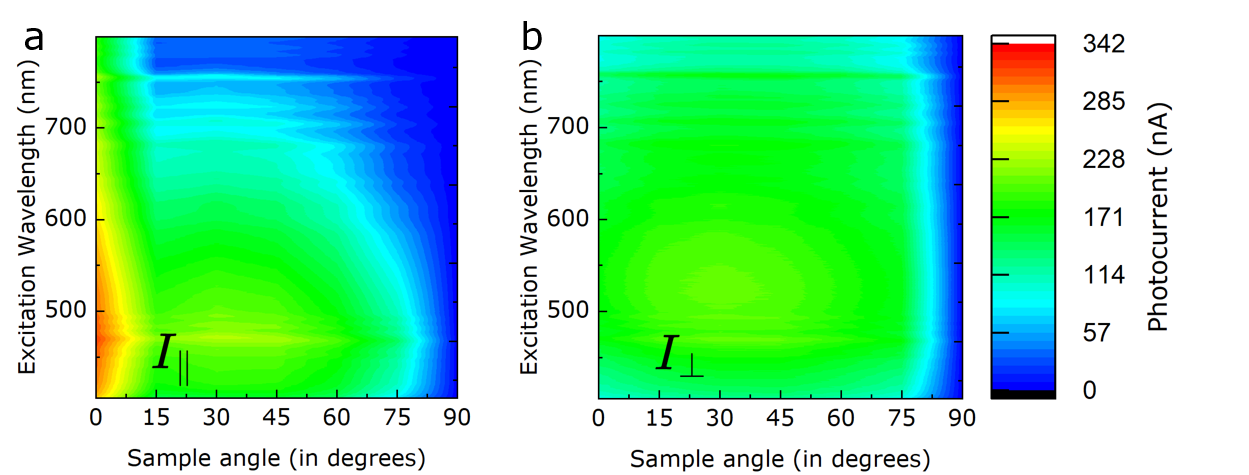}
    \caption{Photocurrent as a function of sample angle and excitation wavelength for two polarization configurations (a) parallel and (b) perpendicular to the principal crystalline-axis.}
    \label{fig:out-of-plane_current}
\end{figure}
\begin{figure}[htbp!]
    \centering
    \includegraphics[width=1\linewidth]{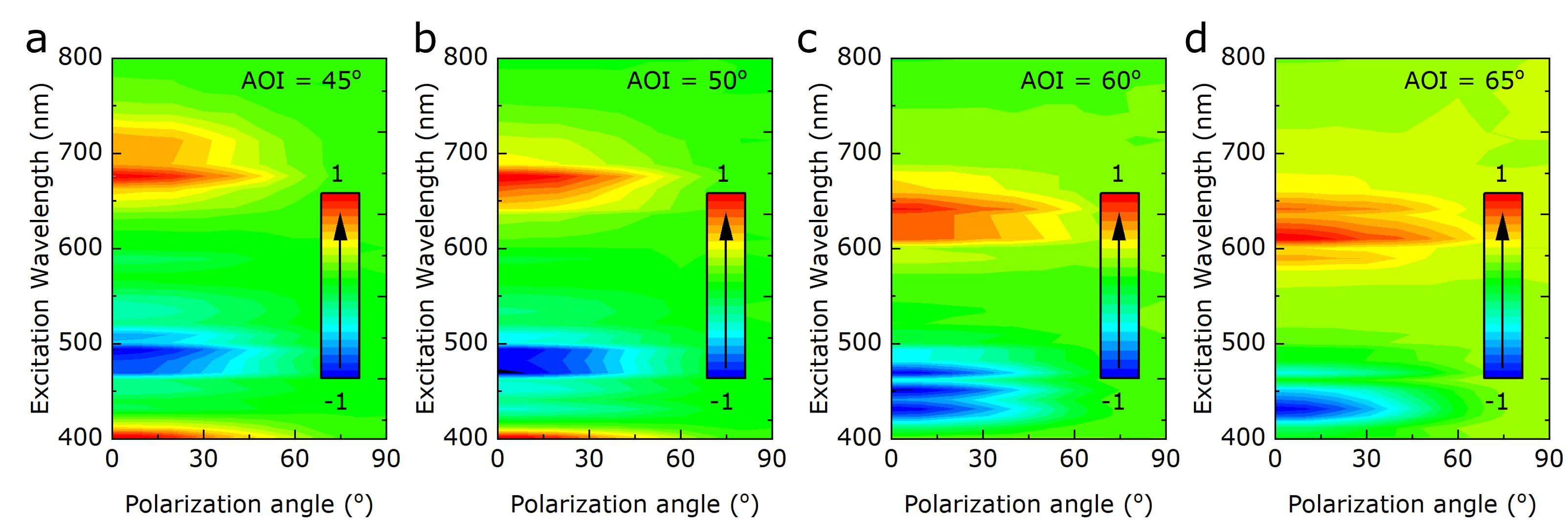}
    \caption{Reflectivity as a function of polarization angle and excitation wavelength for different angles of incidence.}
    \label{fig:reflectivity}
\end{figure}
\clearpage
\subsection*{Responsivity, Detectivity, and Gain Calculations}

The photoresponsivity ($R$) of the device was calculated using
\begin{equation}
R = \frac{I_{\mathrm{ph}}}{P_{\mathrm{device}}},
\end{equation}
where $I_{\mathrm{ph}} = I_{\mathrm{light}} - I_{\mathrm{dark}}$ is the photocurrent and $P_{\mathrm{device}}$ is the optical power incident on the active device area.

The incident optical power on the device was estimated from the measured optical power using a geometric scaling factor,
\begin{equation}
P_{\mathrm{device}} = \left(\frac{P_{\mathrm{measured}}}{A_{\mathrm{sensor}}}\right) A_{\mathrm{device}},
\end{equation}
where $A_{\mathrm{sensor}} = 0.7088~\mathrm{cm}^2$ is the effective sensor aperture area (corresponding to a 0.95 cm diameter aperture), and $A_{\mathrm{device}} = 1483.4~\mathrm{\mu m}^2$ is the active device area.

The specific detectivity ($D^*$) was calculated as
\begin{equation}
D^* = R \sqrt{\frac{A_\mathrm{device}}{2 e I_{\mathrm{dark}}}},
\end{equation}
where $A$ is the active device area, $e$ is the elementary charge, and $I_{\mathrm{dark}}$ is the dark current. For the present device, $I_{\mathrm{dark}} = 21.7~\mathrm{nA}$.

The photocurrent gain ($\Gamma$) was calculated as
\begin{equation}
    \Gamma = h\nu \frac{I_{\mathrm{ph}}}{P\eta e} = R \frac{h\nu}{\eta e}
\end{equation}
For simplicity, we assumed a quantum efficiency of $\eta$ = 1, showing maximum possible detectivity.
\subsection*{Literature Comparison}
\begin{table*}[htbp!]
\centering
\caption{Benchmarking of photodetectors based on 2D magnetic materials.}
\begin{tabular}{lccccc}
\hline
Material & Spectral range (nm) & Bias (V) & Responsivity (A/W) & Detectivity (Jones) & Reference \\
\hline

CrCl$_3$ & 300 - 1700 nm & 3 & $1.85 \times 10^4$ & $8.6 \times 10^{14}$ & This work \\

NiPS$_3$ & 254 - 1020 nm & 0 & $2.3 \times 10^{-3}$ & $6.2 \times 10^9$ & \cite{Zong2024} \\
NiPS$_3$ & 254 nm & 10 & $1.3 \times 10^{-1}$ & $1.2 \times 10^{12}$ & \cite{Chu2017} \\

FePS$_3$ & 363 - 940 nm & 5 & $2.6 \times 10^3$ & -- & \cite{Ramos2021} \\
FePS$_3$ & 254 nm & 0.06 & $0.2$ & -- & \cite{Gao2018} \\

MnPS$_3$ & 365 nm & 10 & $2.9 \times 10^2$ & -- & \cite{Kumar2019} \\

CrSBr & 514 nm & 0 & $2.6 \times 10^{-4}$ & $3.4 \times 10^{8}$ & \cite{Panda2023} \\
CrSBr & 405 - 1550 nm & 0 & $1.1$ V/W & -- & \cite{Zhou2025}\\
\hline
\end{tabular}
\end{table*}
\printbibliography